\documentclass{llncs}

 \usepackage{graphicx}
 \usepackage{epsfig}

\usepackage{amssymb}

\makeatother

\begin{document}
\pagestyle{empty}

\mainmatter

\title{Statistical Economics on Multi-Variable Layered Networks}

\author{Tom Erez\inst{1}
\thanks{
Corresponding Author. Email: etom@pob.huji.ac.il.\protect \\Tom thanks {}``SARA Computing and Networking Services{}'' for generously providing
machine time for the simulations.
} \and
 Martin Hohnisch\inst{2} \thanks{
M.H. gratefully acknowledges financial support from DFG grant TR120/12-1.
} \and
 Sorin Solomon\inst{1} \thanks{
The research of S.S. was supported in part by a grant from the Israeli Science
Foundation.\protect \\
The authors thank Prof. Dietrich Stauffer for advice and help.
}  }

\institute{Racah Institute of Physics\\
Hebrew University, Givat Ram\\
Jerusalem 91904, Israel\\
\and
Research Group Hildenbrand,\\
Department of Economics, University of Bonn,\\
and Department of Mathematics,\\
University of Bielefeld,\\
Germany}

\maketitle
\begin{abstract}
We propose a Statistical-Mechanics inspired framework for modeling economic
systems. Each agent composing the economic system is characterized by a few
variables of distinct nature (e.g. saving ratio, expectations, etc.). The agents
interact locally by their individual variables: for example, people working
in the same office may influence their peers' expectations (optimism/pessimism
are contagious) , while people living in the same neighborhood may influence
their peers' saving patterns (stinginess/largeness are contagious). Thus, for
each type of variable there exists a different underlying social network, which
we refer to as a {}``layer{}''. Each layer connects the same set of agents
by a different set of links defining a different topology. In different layers,
the nature of the variables and their dynamics may be different (Ising, Heisenberg,
matrix models, etc). The different variables belonging to the same agent interact
(the level of optimism of an agent may influence its saving level), thus coupling
the various layers. We present a simple instance of such a network, where a
one-dimensional Ising chain (representing the interaction between the optimist-pessimist
expectations) is coupled through a random site-to-site mapping to a one-dimensional
generalized Blume-Capel chain (representing the dynamics of the agents' saving
ratios). In the absence of coupling between the layers, the one-dimensional
systems describing respectively the expectations and the saving ratios do not
feature any ordered phase (herding). Yet, such a herding phase emerges in the
coupled system, highlighting the non-trivial nature of the present framework. 
\end{abstract}

\section{Introduction}

Statistical Mechanics provides a universal framework to model and analyze large
systems with interacting components, even beyond the traditional realm of physics.
Many techniques of Statistical Mechanics are applied today in diverse areas
of scientific research, such as biology, sociology and economics\cite{stauffer}.
A decade ago, the economist Jean-Michel Grandmont\cite{4} coined the term \emph{Statistical
Economics}, pointing to the analogy between large economic systems and large
physical systems, and suggested that large economic systems can be appropriately
modeled using techniques developed in Statistical Mechanics\cite{brock}. 

For Statistical Economics to benefit from the vast knowledge that accumulated
in Statistical Mechanics, it is crucial to establish which structures are shared
by physical and economic systems, and which are essentially different. Economic
systems, like physical systems, involve local interactions of both attractive
and repulsive type. To formalize such interactions, a Boltzmann-Gibbs measure
derived from a Hamiltonian function is an appropriate formal object. However,
a proper interpretation of the Hamiltonian is still an open issue in economics.
While in the present paper we work within the Hamiltonian formalism, the ideas
herein apply also to dynamical systems\cite{Lotka,AB-model} that do not admit
an energy function, i.e. the interactions may be asymmetric. 

The locality of the interactions raises the issue of the structure of the underlying
network. In physics, a regular lattice is often an adequate approximation of
the graph structure underlying the interaction in the real world. In contrast,
economic dynamics involve social networks, for which a regular graph topology
is usually a very rough and often unjustifiable approximation. Scale-free graphs
are presumably more appropriate, yet empirical evidence on the actual structure
of communication and social networks is only beginning to emerge\cite{Andrzej_Nowak}.
For theoretical and empirical results in this emerging field we refer the reader
to \cite{networks,Havlin}. 

Lately agent-based economic models have been introduced that take into account
the inhomogeneous character of the various agent variables . However, the connections/
interactions between the agents were usually represented by a single network.
In the present paper we address a different aspect of economic systems: their
multi-layered structure (Figure 1). 

Isolating the economic interactions of one type (e.g. wealth exchange) from
the wider economic context of the agents' interactions (e.g. exchange of convictions,
preferences, expectations, risk attitudes, assets etc.) is often unrealistic.
In general, individual economic decisions result from an interplay of such distinct
variables. In fact, it is a pillar of classical economics that diverse individual
variables interrelatedly determine decisions. Yet, such systems have found so
far little attention in Statistical Economics models\footnote{%
Multi-layered models of a different nature were considered in other areas of
physics-inspired research, like neural networks\cite{NN} and opinion formation\cite{snajd}.
}. 

In the present paper we aim at formulating a general framework for
modeling economic systems, where each agent is characterized
by multiple variables of distinct nature. Each type of variable
constitutes a layer with a particular graph structure, with only
the set of nodes (agents) common to all layers. In addition to
local interaction of neighboring variables within a layer,
variables associated with the same agent in different layers
interact. This introduces coupling between the layers, which qualitatively
changes the dynamical pattern of every layer. We call a system with
the above properties a ``Solomon Network'' (SN) \footnote{This nomenclature
has been introduced by Dietrich Stauffer with reference to the biblical dilemma of an agent being ``claimed'' by (i.e. is part of) different social networks.}.

At first glance, the dynamics of each layer in a SN may look similar to a single
variable system in some external field provided by the rest of the variables.
Yet, SN models are unique in the sense that the interaction between layers is
bi-directional, allowing for feedback between each layer and its {}``external{}''
field. The dramatic effect that this feedback can have on the system behavior
is demonstrated already at the level of the simplest model described below.
This is a step toward substituting the usual closed system and stationary/quenched
environment paradigm with economics models that provide a wider and interactive
(and reactive) representation of the environment. 

The topology of the coupling between the layers in a SN is formally characterized
by the mapping between the nodes of the different layers. In physical systems
the graph structure is often formalizing physical proximity, and therefore if
a physical system would be under investigation, we would expect such a mapping
to be isometric (preserving distance). In contrast, social networks formalize
communication links rather than physical proximity. Therefore, preserving distance
is not a natural property of a mapping between different social networks. It
may well be the case that an agent's reference group for a certain social or
economic interaction is entirely different from her reference group for another
type of interaction. For example, we may think of a scenario where, for certain
people, consumption and saving patterns are formed by comparing themselves to
people who live in their residential neighborhood, while their expectation of
the general state of the market is formed through discussions with their colleagues
at work. These two reference groups may be totally distinct, or have any degree
of correlation (overlap). In general, the graph structures associated with different
layers may differ: for example, a random graph may be coupled with a regular
graph. Moreover, one may extend the above definition of SN to include cases
in which the site correspondence between the layers is not one-to-one (e.g.
people that do not go to work, or people who work in more then one place). 

As a first step toward an understanding of such models and the properties of
corresponding economic systems, we formulate in Section 2 a simple model with
only two layers. In section 3 we discuss the theoretically expected properties
of the model and in section 4 we study the model by using Monte Carlo simulations.
We return to the general paradigm and discuss its possible ramifications in
Section 5.

\section{The model}
Let $A$ denote a set consisting of $N$ economic agents. Each agent
$i \in A$ is characterized by two random variables. First, the
variable $X_i$ represents the agent's saving ratio, i.e. the
proportion of the individual budget which is saved, rather than
consumed. Each variable $X_i$ takes values in the set
$\mathcal{S}_x=\{1,2, \dots ,Q\}$ with some natural number $Q$. To
interpret $X_i$ as a savings ratio, one considers $\frac{1}{Q}X_i$
rather than $X_i$.

Second, the variable $S_j$ represents the agent's
expectation about the prospects of the economy. For simplicity, we allow only two
individual states, called ``optimism'' and ``pessimism''.
Accordingly, each variable $S_j$ takes values in the set
$\mathcal{S}_s =\{-1,1\}$, and we arbitrarily interpret $1$ as
``optimism'', and $-1$ as ``pessimism''.

Thus we define two layers for the model consisting respectively of the 
variables $(X_i)_{i \in A}$ and $(S_j)_{j \in A}$. Both layers are
modeled as spin systems with nearest-neighbor interaction, with
respect to the graph structure underlying each layer. In addition,
there is interaction between those pairs of variables from
different layers, which correspond to the same agent. 
Thereby the two spin-fields are coupled.

In the present simple SN model,  we confine
ourselves to one-dimensional layers with periodic boundary conditions
(chains). Thus, the agent $i$'s neighbors in the
$X$-layer  are $(i+1)$mod$N$ and $(i-1)$mod$N$. The neighborhood
relation on the $S$-layer is specified through a permutation $r$ of the
elements of $A$. More precisely, the agent $i$ is represented 
on the layer $S$ by the site $r(i)$.
Accordingly, the two variables associated with agent $i\in A$ are
$X_i$ and $S_{r(i)}$.
The neighbors of $r(i)$ in $S$ are $(r(i)+1)$mod$N$ and $(r(i)-1)$mod$N$.
Thus the agent $i$ interacts directly with four agents: 
$i+1$, $i-1$, $r^{-1} (r(i) +1)$, $r^{-1} (r(i)-1)$.
Note that the interaction with the first pair is of a different nature 
(and in fact parametrized by different variables) 
than the interaction with the other pair. 
Note also that in this model we have 
chosen a one-to-one, totally random correspondence between the layers.

We  denote by $\mathcal{S}=\{-1,1\} \times \{1,2, \dots ,Q \}$ the
individual spin-space of the composite system. A configuration
assigns to each agent $i \in A$ values  from $\mathcal{S}$. The
configuration space, denoted as usual by $\mathcal{S}^A$, consists
of $(2Q)^N$ configurations.

The dynamics is defined by introducing 
the Hamiltonian function $H$ that assigns to each configuration 
$\omega \in \mathcal{S}^A$ some
real number $H(\omega)$, such that the probability of $\omega$ is
given by the Boltzmann distribution associated with $H$
\begin{equation}
Pr(\omega)=1/Z_T\exp(-H(\omega)/T).
\end{equation}
 with $T$ denoting the temperature parameter of the Boltzmann distribution,
and $Z_T$ being the normalizing constant given by
$Z_T=\sum_{\omega \in \mathcal{S}^A}\exp(-H(\omega)/T)$.
The ratio $H/T$ measures the relative  strength 
of the social interaction of the system against other random perturbations.
For $T=0$  the peer pressure is absolute,
and basically the individual has no freedom. 
for $T=\infty$, 
there is no influence between any of the variables and all states of the
system are equally probable.
 
In our model, there are three components of interaction:
1) the nearest-neighbor interaction among different agents on the $X$-layer,
2) the nearest-neighbor interaction among different agents on the
$S$-layer, and 3) interaction between the
variables corresponding to the same agent in the two layers.

The term in the Hamiltonian representing the interaction within the $X$-layer is
specified as
\begin{equation}
H_X(\omega)=J_x \sum_{<i,j>_X}(X_i(\omega) - X_j(\omega))^2.
\label{hx}
\end{equation}
with $<i,j>_X$ denoting the summation over pairs of nearest neighbors
according to the $X$-layer topology.
 The basic economic content of this specification is the notion of
herding behavior \cite{HB,herd}. That notion stands in economics for a
variety of effects where market actors, such as stock traders,
consumers, managers etc., tend to align their opinions and/or
actions with those in their social neighborhood. In our case,
if the current state $X_i$ of the agent $i$ is very different from the 
typical opinions $X_j$ of its neighbors $j$, Eq. \ref{hx} indicates
a strong energy preference for the next moves to bring $X_i$ and $X_j$ closer.
This specification is in accordance with
experimental results from social psychology \cite{Andrzej_Nowak,LF,Latane} 
suggesting that the likelihood of realignment of one's own actions and
beliefs (i.e. the saving behavior, in the present model) with
those observed in the reference group, will increase with the
perceived difference in those actions and beliefs. In physics, a
Hamiltonian of the above type have been used, for instance, in the
Blume-Capel model\cite{blume,capel}.

A similar herding behavior is observed for the agents' expectations:  
the expectations of the individuals, with respect to the
future market behavior, will tend to conform with the opinion 
of their social surrounding. 
Thus the interaction between the agents' expectations has a 
similar form as Eq. \ref{hx}.
Since we use a rougher characterization of
expectations ($-1$ corresponds to a pessimist view, and $+1$
corresponds to an optimist view), the interaction on the
$S$-layer reduces to the classical Ising-model Hamiltonian:
\begin{equation}
H_S(\omega)=J_s \sum_{<i,j>_S}(S_i(\omega) - S_j(\omega))^2.
\end{equation}
 If $Q=2$, the $X$-layer is also an Ising chain, and the model reduces 
to a classical Ising model on a peculiar SN graph structure. 
This special case was discussed in \cite{M}. 
However, a key property of the SN framework is, in our
view, the distinct character of the respective layers, both in
terms of the spin-space and the underlying graph structure. 

To complete the definition of the system, one has to specify the
interaction between variables
corresponding to the same agent in the two layers, i.e. $X_i$ and
$S_{r(i)}$. This interaction expresses the prominent economic behavioral
regularity that an optimistic agent is likely to save less than a
pessimistic agent\cite{davidson}. Thus we introduce a term that couples the
variables corresponding to the same agent:
\begin{equation}
H_{C}(\omega)=J_c \sum_{i}(X_i(\omega) - C(S_{r(i)}(\omega)))^2.
\end{equation}
The function $C(\cdot)$ goes from $\mathcal{S}_s$ to the real
numbers, with $C(-1)>C(1)$. Obviously, when $J_c=0$, the system breaks down to two
independent chains.

Finally, the Hamiltonian of the system is obtained by adding the three components:
\begin{equation}
H( \omega)= H_X(\omega)+H_S(\omega)+H_C(\omega).
\end{equation}
Since (cf. Eq. 1) the parameters of the model appear dynamically in the 
combination $H/T$, we will often consider the coupling parameters $J$
as constants and study the behavior of the system under the variation of $T$.
Thus in the following we use the parameter $\beta=1/T$ as a measure 
of interaction-to-noise ratio in the economic system.

In the next two sections we will discuss this system
from the theoretical and respectively simulation point of view.

\section{Discussion of the Expected Model Properties}

For complete characterization of the phases of the model, 
one may have to perform numerical simulations 
of the type we present in the next section.  
However, the experience accumulated in the last 30 years
in Statistical Mechanics can be used to extract quite 
a number of predictions and intuitions about the model,
and especially its herding (order-disorder transition) properties.

While the SN model was not studied before, 
the properties of each layer in separate are covered by 
rigorous general theorems \cite{Stanley}. 
To understand the import of the theorems let 
us start with the behavior of the decoupled layers at T=0.
In this case, both layers are in a ``ground state'', i.e.
the state of minimal energy. 

The case of the $S$ layer is easier to analyze.
The minimal energy is realized for $S_i = S_0 \ \forall i$,
with 
$ S_0 = \pm 1$.
In general, for dimensions larger then one,
there exists a {\bf critical value} $T=T_c$ such that for the entire range
$T < T_c$ the system remains ordered.
Indeed, for  $T_c > T > 0$, the configurations are not exactly uniform,
but the $S_i$'s across the entire system remain ``herded'' 
around one of the values $+1$ or $-1$. The degree of herding
of each configuration $\omega$ is measured by the magnetization  
$M$ defined
as the average of the $S_i(\omega)$ over the
entire range $i \in A$:

\begin{equation}
M = \sum_i S_i /N
\end{equation}
The transition at $T_c$ between the ordered ($M\not = 0$)
and disordered ($M=0$) phases
is called , following the Statistical Mechanics conventions, a 
Phase Transition.  
In the present article we will use interchangeably
the words ``magnetization'' and ``herding'' to denote $M$.

Interestingly enough, in one-dimensional systems the order-disorder
phase transition is absent ($T_c = 0$)\cite{Stanley}.
Thus, theory predicts the absence of herding in the absence of coupling 
between the ``expectations'' layer and the ``savings/consumption'' layer
of our model.
In fact, rigorous theorems insure that ANY system 
with local interactions in one dimension is in the 
disordered phase for any $T> 0$.
Thus, even the coupled system will not present herding, as long as the 
two layers have a roughly similar metric 
(i.e. $|r(i)-i|<<N \ \forall i$). 
Only when the two layers have significantly different neighborhood
assignments a herding phase can emerge at finite $T$.
This is because such a system would not be amenable to the format of
a one-dimensional model with local interactions.
Moreover, the transition between the herding and disordered phase
will have a character that is not necessarily the one characteristic 
to any of the higher dimension Ising (or Blume -Capel) models.
The Mote Carlo 
simulations we report in the next section confirm these predictions.

In higher dimensions each of the layers, when taken independently, may 
display quite different critical parameters $T_c^X \not = T_c^S$.  
As one brings the layers ``in contact'' by making them interact, 
the composite system might preserve two independent phase transitions
(if the inter-layer interaction is weak) or change its behavior qualitatively. 

One may ask how relevant is the phase transition analysis for an actual system.
After all, it is {\it a priori} probable that the parameters of the system
realized in nature are far from the critical ones. Thus the system  
would spend its entire life span in just one phase and ignore the other.
The answer is that it falls upon the shoulders of the modeler 
to identify the interactions and parameters that DO matter qualitatively in the
real system. Modeling interactions that do not matter is of course useless
(nobody would model the current financial markets in terms 
of the issue of whether the dollar is or not going to be devaluated by 100\%
during the current trading day
or whether the side of the road on which the cars drive would spontaneously flip).
Thus pessimism and optimism, and in general the values that the
dynamical variables can take, should be and are usually chosen such as to
 reflect operative possibilities
so that the system may very well be in either phase.

Given that a phase transition (say herding-disorder) exists,
one may wonder whether establishing its type (universality class)
in detail is of interest in economics. 
The difference between a discontinuous (first order) phase transition 
and a continuous one may give precious indications on whether
to expect large fluctuations or rather a plain collapse. 
Even if the transition is continuous, it is often important to 
know which of the layers has the dominating role in triggering it.

\section{Monte Carlo Study of the Model}

We simulated the above-defined model using 
 standard Monte Carlo simulation methods\cite{Metropolis}.
In particular, we update the agents' state in a random order.
Each Monte Carlo Step (MCS) is composed of $N$ such random
selections. In each computational step, after an agent $i$ is
chosen at random, we run a Heat-Bath algorithm
\cite{Binder-Landau} to generate the next state. The new values
for the selected $X_i$ and $S_i$ are sampled according to the Boltzmann
distribution (Eq. 1) with respect to the potential $H$ (cf. Eq. 5): given a
particular configuration of the system $\omega \in \mathcal{S}^A$,
let us denote $\omega^i_{x,s}$ the configuration of the system
which is identical to $\omega$, except for the variables associated with agent $i$,
where $X_i=x$ and $S_{r(i)}=s$. The probability to move from
 $\omega$ to $\omega_{x,s}$ is given by the following equation:
\begin{equation}
P(x,s)=1/Z^i_T\exp(-H(\omega^i_{x,s})/T)
\end{equation}
where $Z^i_T=\sum_{(x,s) \in \mathcal{S}}
\exp(-H(\omega^i_{x,s})/T)$ i.e. the sum over all
possible values for $X_i$ and $S_{r(i)}$.

In the present investigation, we simulated a system with $N=10^6$
agents and $Q=10$ saving ratios. Our simulations around the
$T_c$ consisted of $6.4*10^5$ MCS, and the results
were averaged over the last $3.2*10^5$ MCS. Less iterations were
needed for the $T$ ranges outside the critical
slowing-down zone, where $2*10^4$ MCS were sufficient, and the
results were averaged over the last $10^4$ MCS in that case.
Unless stated otherwise, simulations were run with
$J_s=J_x=J_c=\frac{1}{3}$, $C(-1)=8$ and $C(1)=3$. The initial
conditions were set to either the ordered configuration 
($S_j=-1$ and $X_i=8$) or to a totally random one.
 We obtained the randomly generated permutations $r$ by 
mapping every $X$-layer site to a $S$-layer site 
chosen from the entire lattice with equal probability.

We found a phase transition occurring at a critical value of
approximately $T_c=2.79$. Both layers become ordered below
$T_c$. Figures 2 and 3 depict the herding of each of the two
layer as a function of $T$. The functional dependence
of the herding $M$ on $T$ close to $T_c$ appears to be: (cf. Figures 2b and 3b)
\begin{equation}
M \propto 1/log (T_c - T).
\end{equation}
In contrast, \cite{M} measured the critical exponent 
$\beta$ of the SN model with two Ising layers to be $1/2$ 
(i.e. in the Ising universality class). 
Therefore this feature of the model is attributed to the coupling of two layers carrying
 different dynamics, 
and not merely a result of the particular topology.

The two ordered states ($M<0$ and $M>0$) for $T<T_c$
can be called ``pessimistic economy'' and
``optimistic economy'' respectively. The transition between them is 
a first order one similar to the transition when crossing
the line of zero magnetic field $H=0$ 
below the critical Ising temperature.
Thus our model is capable of capturing empirical results describing dramatic swings in 
the market mood\cite{swings}.
Moreover the model predicts that in the herding phase, the transition
between an optimistic and pessimistic mood in the $S$-layer induces 
a transition between the ``saving'' and ``spending'' modes in the $X$-layer.

A variable of major interest for economics is the empirical
distribution of the $X$-layer, since it characterizes
 the aggregate saving in the economy. Figure 4 depict empirical
distributions of the $X$-layer at different $T$'s. The
distribution is symmetric above $T_c$, i.e. in
the disordered state. Below $T_c$, the distribution is
skewed, reflecting symmetry breaking in the herded state. If
$C(-1)$ and $C(1)$ are further apart, the empirical distribution of
the $X$-layer for $T>T_c$ becomes bimodal (Figure 5).

In accordance with the theoretical prediction,
no phase transition was found in simulations
where $r(\cdot)$ was the identity permutation (results not shown), 
since the system is essentially one-dimensional. 
Also, when the layers are defined on similar lattices 
(as is the case in our simulations), 
one may construct the $r$-transformation to be bound 
(i.e. $ |r(i)-i|< C_0 \ \forall i $).
In this case as well, no ordered phase emerges
(results not shown).
According to the theory, the ordered phase shrinks to $T=0$
as one lowers $J_c \rightarrow 0$ (Figure 6).

In order to connect the model to the actual economic reality,
one can calibrate the model by comparing it to empirically 
measurable data.
In particular one can measure the size of the 
various groups as defined by their expectations and saving patterns.
From this, one can infer by inverse engineering the realistic
values for the coupling constants in the model.
We studied the relation between the sizes of the groups 
and the values of the $J$'s as described below.
In the ``pessimistic'' phase, the agents
$i\in A$ with $s_{r(i)}=-1$ and $x_i\in\{6,\dots,10\}$ constitutes the
majority. In contrast, we can arbitrarily define three different
minority groups. First, the set of agents with $s_{r(i)}=1$  and
$x_i\in\{6,\dots,10\}$ is called $S$-minority; second, the set of agents with $s_{r(i)}=-1$ and
$x_i\in\{1,\dots,5\}$ is called $X$-minority; and the set of agents with $s_{r(i)}=1$ and
$x_i\in\{1,\dots,5\}$ is called $SX$-minority. The values chosen for the coupling constants $J$ 
 determine the balance between the different minority
groups. Figure 7 shows the distribution of the different minority
groups for various ratios of the interaction constants.
 If a certain interaction constant increases, 
the proportion of the corresponding minority group
 will decline for every $T<T_c$. For
example, when $J_s=0.6$ and $J_x=0.2$, the $X$-minority group is
downsized considerably (Figure 7). This is because every agent
``prefers'' conforming to its neighbors on the $S$-layer (i.e.
having the same $S$ value), even at the ``price'' of non-conformity
with its neighbors on the $X$-layer (i.e. having a different $X$
value). Also, this figure shows certain symmetry of the J's. Both
layers respond equally to increased preference, at least in the
aspect of minority ratios - the same preference imbalance results
in the same bias in the distribution between minority groups.

\section{Conclusions and Outlook}

In the present paper we have presented one of the simplest instances
of an application of the SN framework to economic systems modeling. 
Already at this level one obtains interesting 
results, underlying the importance of the coupling of the various
types of economic variables in determining the order-disorder dynamics of
the global system.

Using representative agents, population averages or differential equations
governing the global (macro-economic) system variables,
one might have guessed that by coupling weakly two types of economic 
variables that do not present herding the resulting system will be in the 
disordered phase, i.e. the coupled system would behave 
somewhat as the average of the 
respective behaviors of its components.
Instead, using theoretical and simulation tools adapted from Statistical Mechanics 
one finds that the coupling of two disordered economic variables 
has a dramatic herding effect on both variables. 
This shows the importance of the further study of SN
models in uncovering the subtle effects of feedback 
in systems with multiple economic variables.

The study of SN models can be developed in quite a few directions.
First, instead of the completely random mapping $r(i)$ between the
positions of the nodes in different layers, one may consider 
less extreme alternatives: after all, there is often some correlation between 
the work place of people and their residence districts.
For instance, one may consider mappings in which for a significant number of of
$i$'s one has $r(i)=i$, or mappings which preserve the topology
of the neighborhoods but not their metric.
  This might elucidate which properties of the SN model
 can be traced to the mere coupling between layers 
(and thus possibly detectable by standard macro-economic methods), 
and which are a result of the particular
inter-layer projection afforded by the SN framework.

Second, one should consider inhomogeneous layers, e.g. scale-free 
or small-world, and study projections $r(i)$ of different tyes (e.g. preserving 
rank or not etc.). Of course the various layers may differ in
their properties, such as their scaling exponent, clustering coefficient
or K-core structure\cite{VISH}.

The SN modeling framework can also be extended to include global
feedback schemes, which are particularly relevant in economic
applications\cite{Bornholdt}. We intend to pursue this direction of research
elsewhere\cite{new}.

To make contact with reality, one should find and use the parameter values 
that represent faithfully and realistically the empirics of the modeled 
economic phenomenon.
 For example, by looking at relative sizes of the herding groups 
with respect to various variables, 
one may receive hints as to the relative strength of the couplings 
governing each variable.
This would require the use of cross-correlational micro-data. 
Such data exists,
for instance, for savings and consumer confidence\cite{microdata}.

In summary, we believe that the framework of Solomon Networks
addresses a key property of large economic systems - the
cross-interaction between the dynamics of economic variables of different types.
 Further investigation of such models may shed light on the mechanisms governing economic
dynamics, and increase our understanding of the complex system
called Economy.

\begin{figure}[htp1]
\begin{center}
\includegraphics[angle=0,scale=0.6]{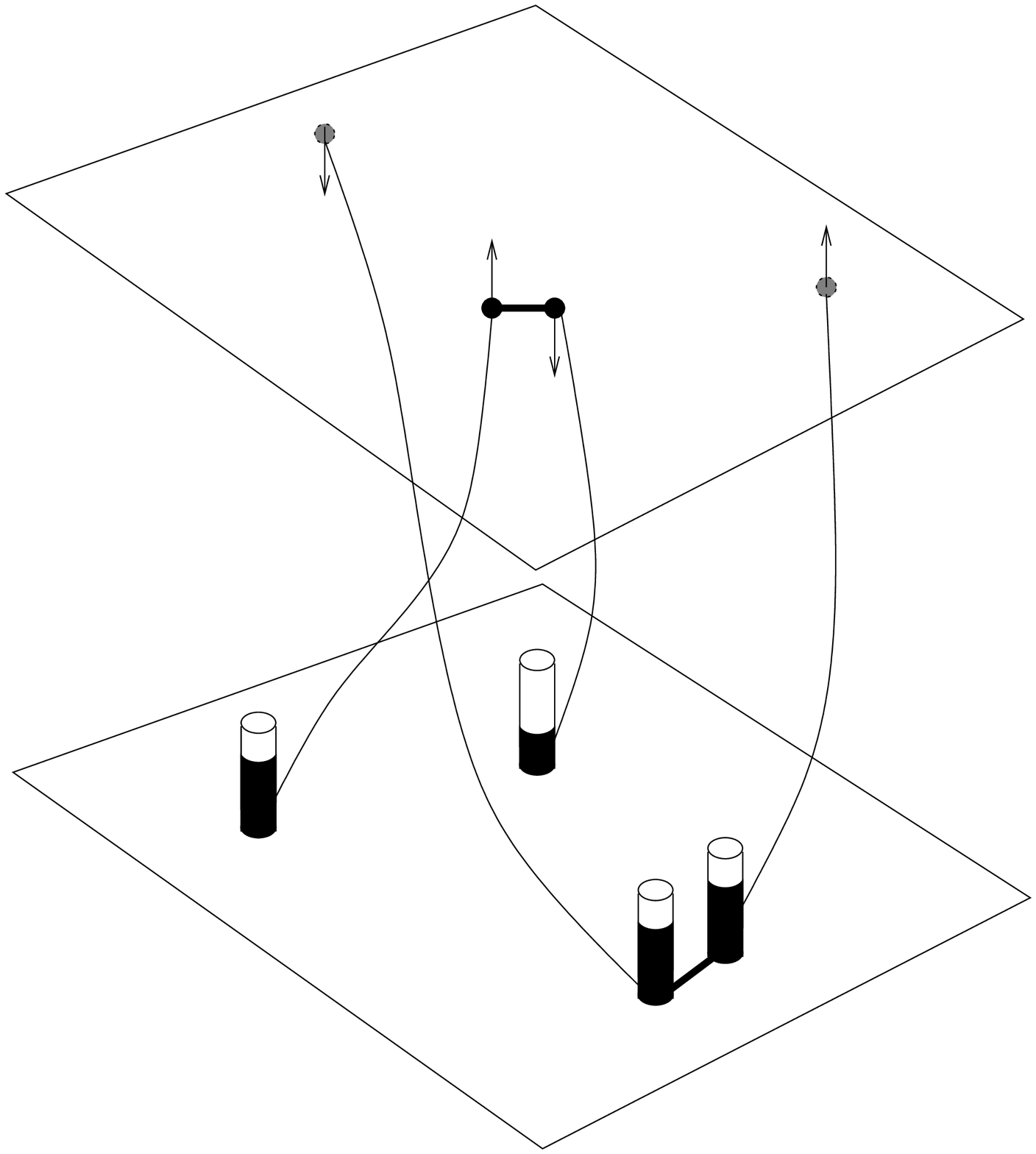}\\
\caption{A schematic representation of the Solomon Network architecture. Each agent is represented by two variables, one on each layer, connected by a curved line, while thick lines represent nearest-neighbor relations on each layer. Not all the intra-layer neighbors of the agents are depicted. Two elements are of special interest: the variables on each layer are of a different type, and the neighbors of an agent on one layer are not her neighbors on the other layer.}
\end{center}
\end{figure}

\begin{figure}[htp]
\begin{center}
\includegraphics[angle=-90,scale=0.4]{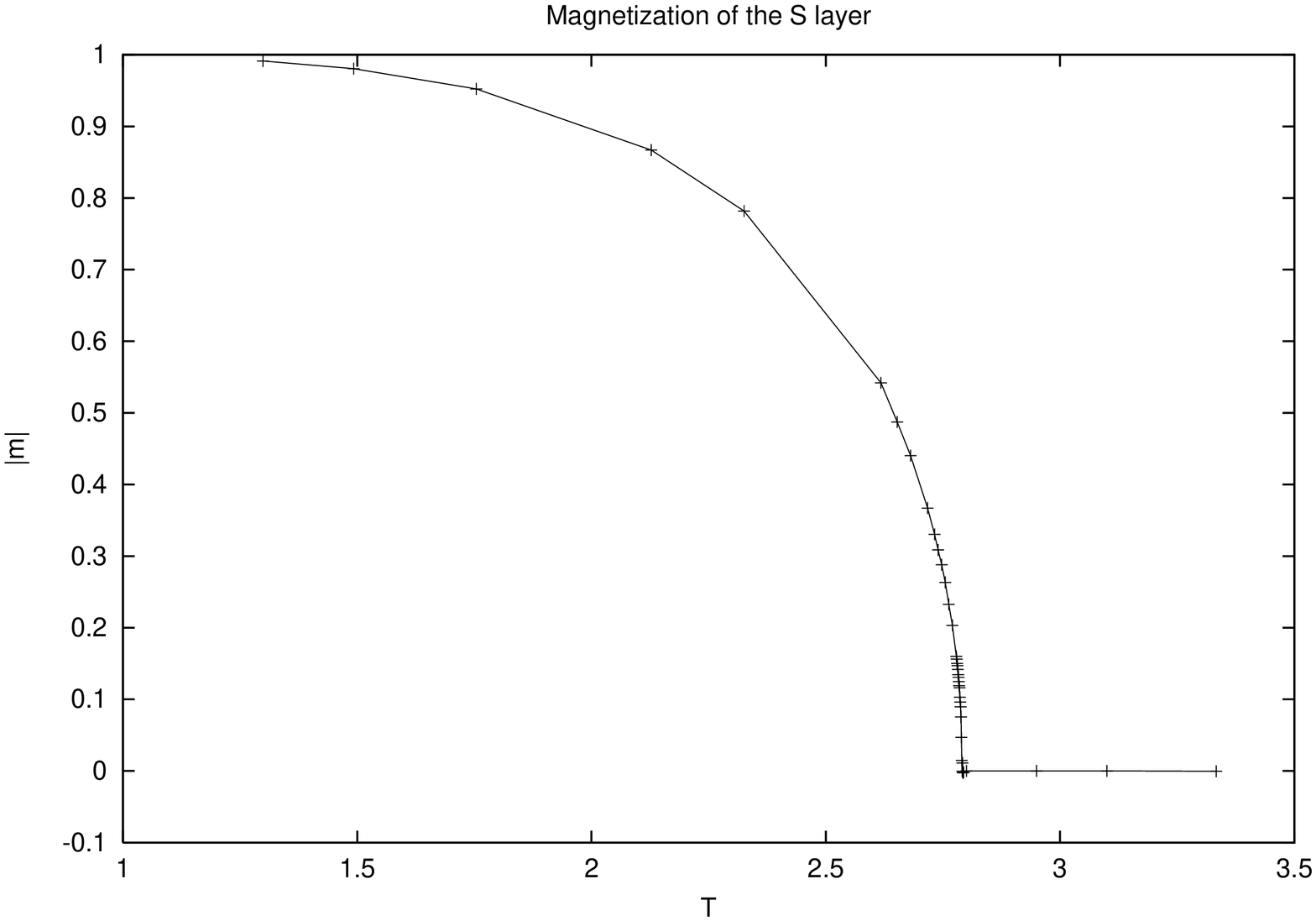}\\
\includegraphics [angle=-90,scale=0.4]{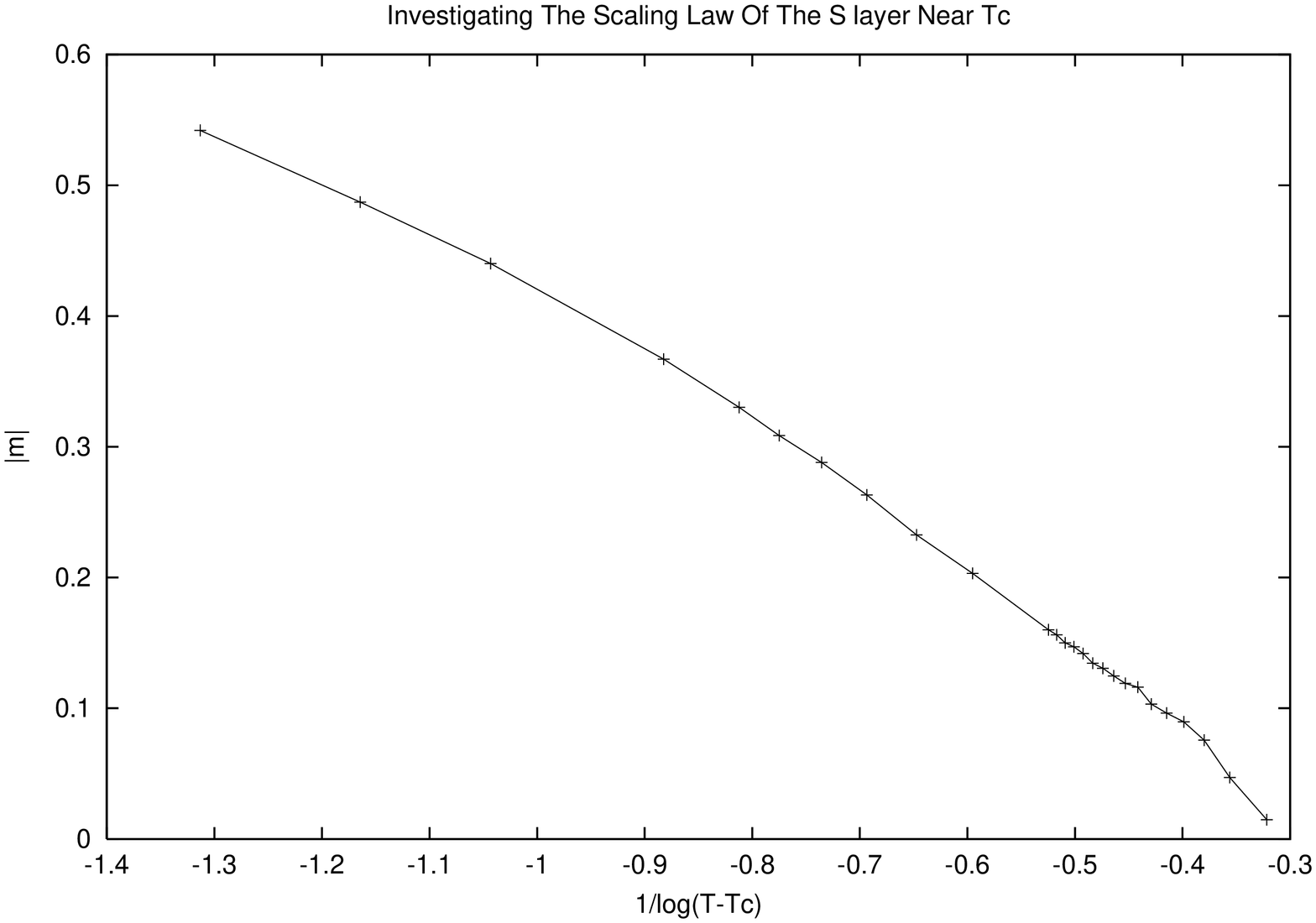}
\caption{Herding of the  expectations as a function of T. 
Figure 2a dispalys the main feature of the model:
as opposed to the behavior of the separate expectations and
savings layers, the composite system present 2 distinct phases:
one without herding $M=0$ for $T > Tc$ 
and one with $M>0$ for $T < T_c$
Figure
2b demonstartes the dramatic nature of the vanishing of the herding $M$ 
of the expectations
as one approches $T_c$ by showing that $M \propto 1/ log(T_c -T)$, i.e. all the derivatives are infinite at the vanishing point.}
\end{center}
\end{figure}

\begin{figure}[htp2]
\begin{center}
\includegraphics[angle=-90,scale=0.4]{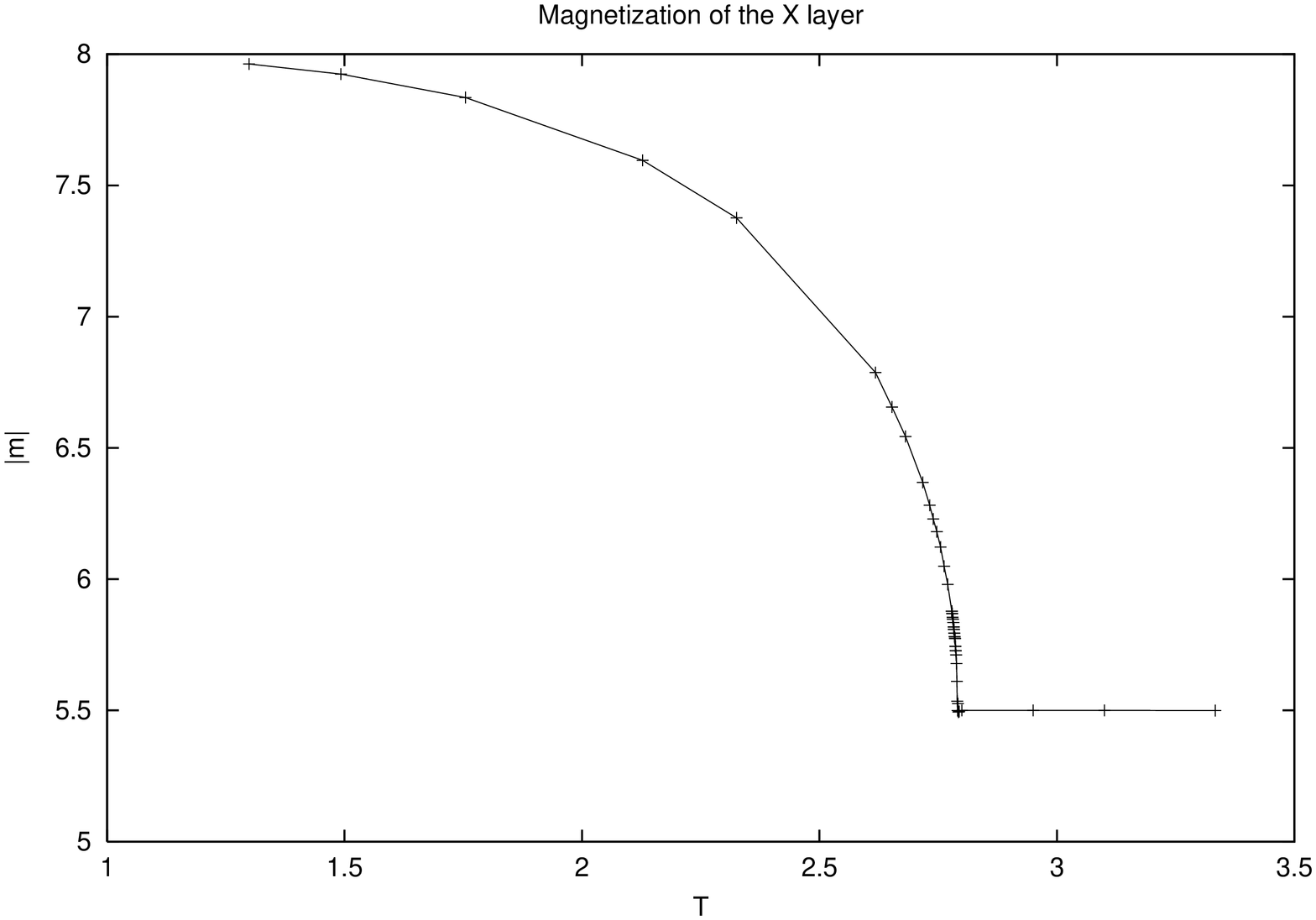}\\
\includegraphics [angle=-90,scale=0.4]{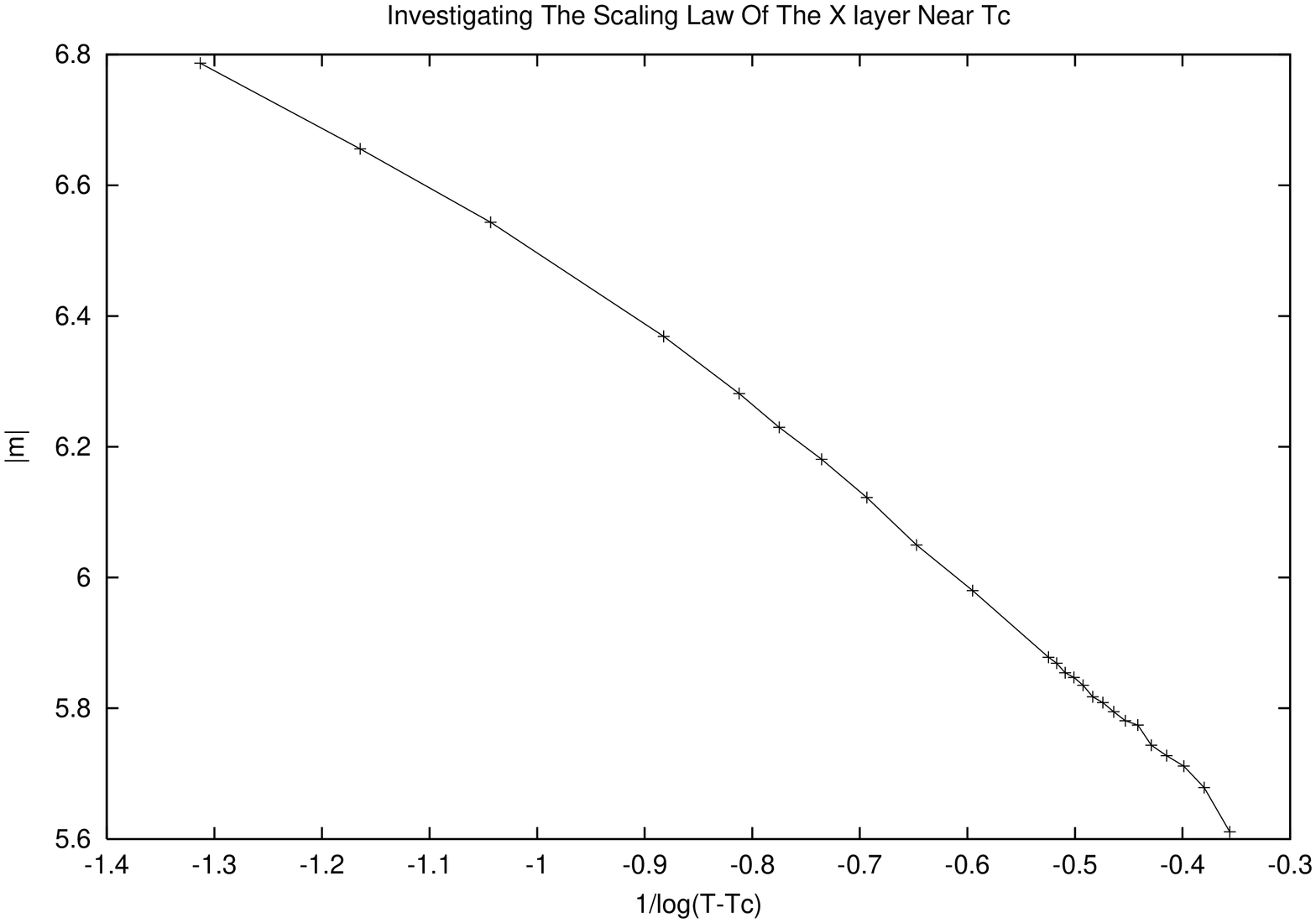}
\caption{Similar results as in Figure 2, this time for the herding of the savings ratios. 
Since $X_i$ ranges between
1 and 10, 5.5 is the average value in the unordered phase.  }
\end{center}
\end{figure}

\begin{figure}[htp3]
\begin{center}
\includegraphics[angle=-90,scale=0.35]{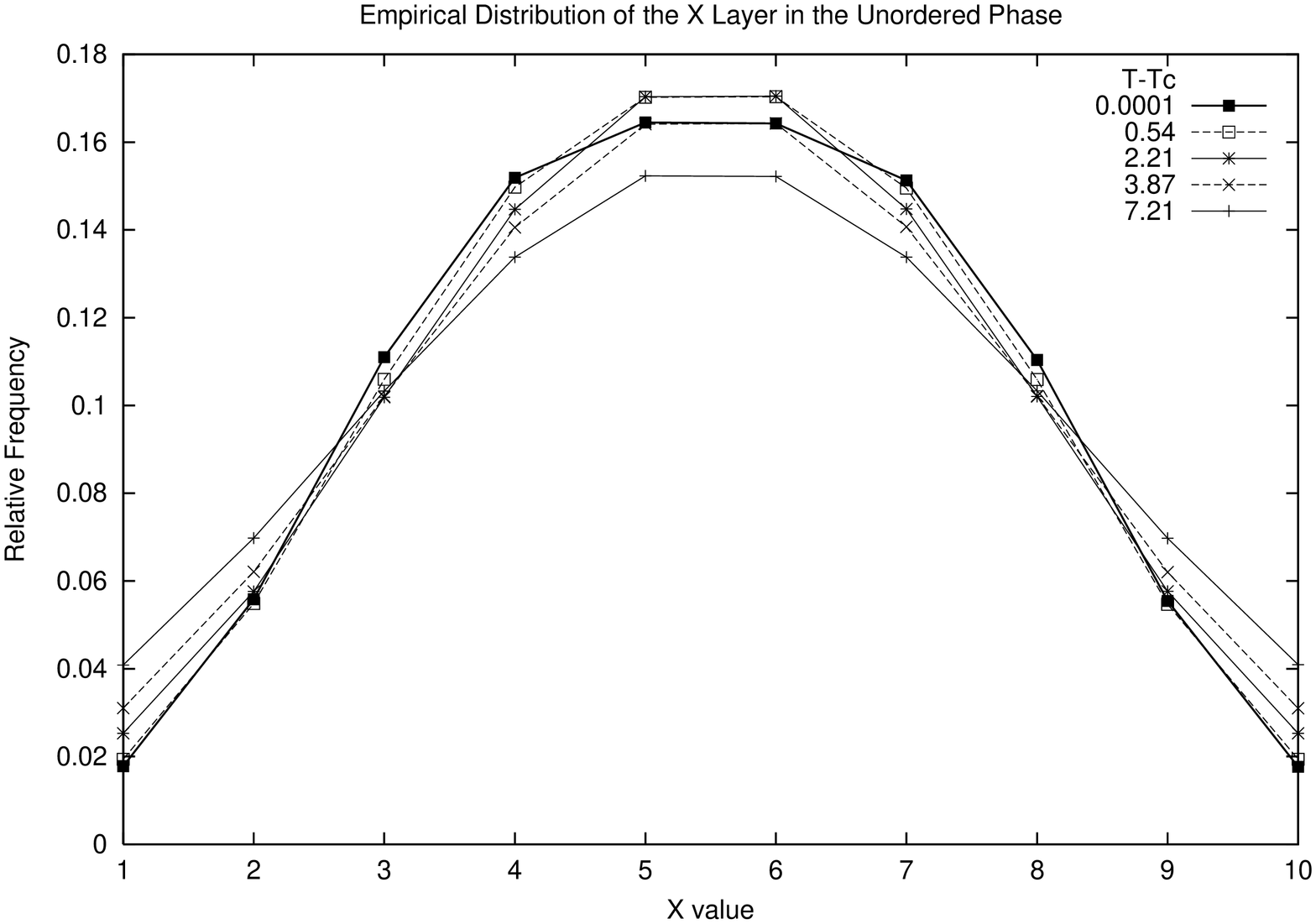}\\
\includegraphics [angle=-90,scale=0.35]{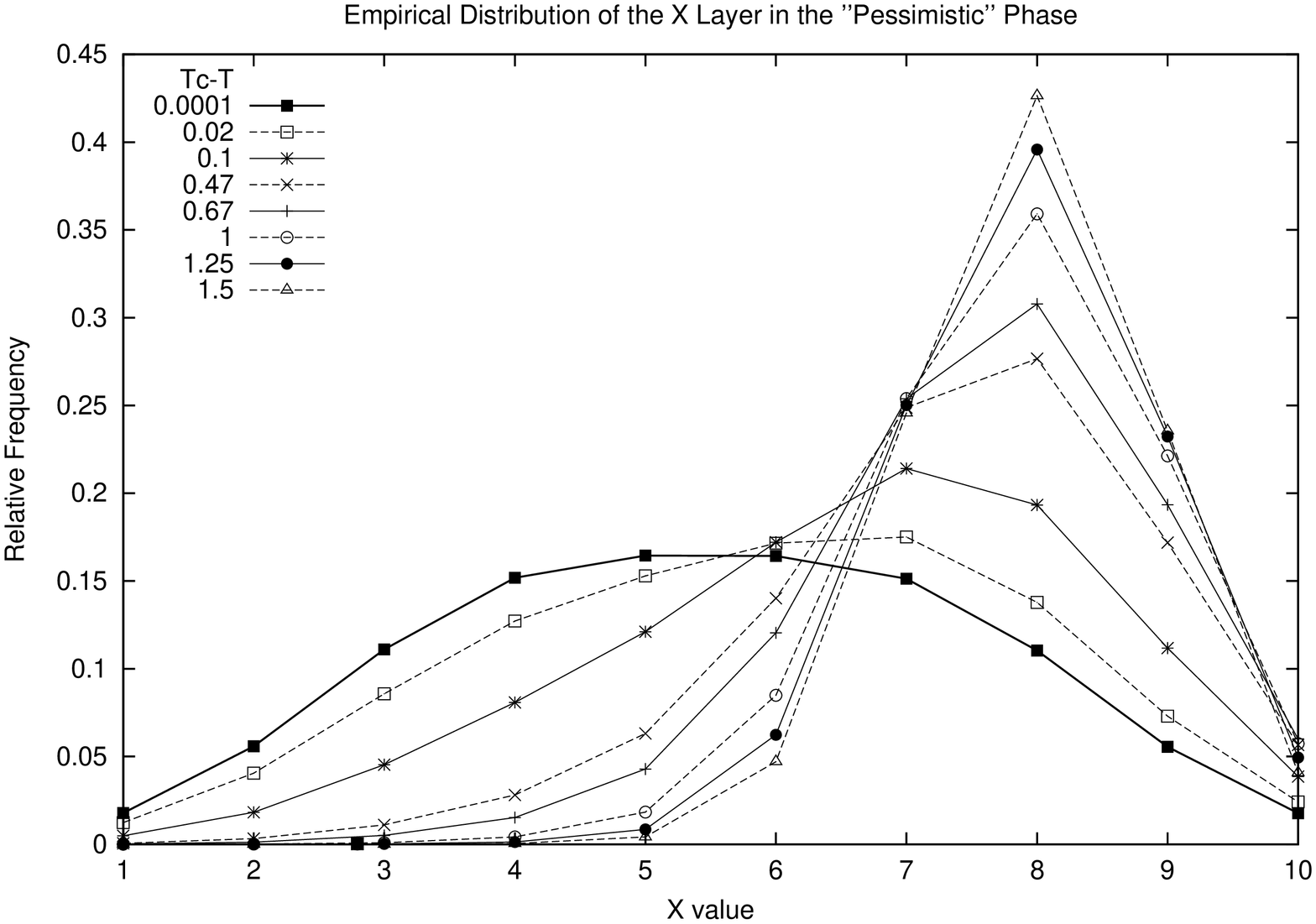}\\
\caption{Empirical distribution of individual saving ratios ($X_i$) at
different values of T. 
Figure 4a depicts the distribution of $X_i$'s at various $T$'s in
the non-herding phase. At $T=\infty$, the distribution is uniform
(not shown). Note how all distributions are symmetric in the non-herding phase. 
Figure
4b depicts the empirical distribution of $X_i$ 
at various values of $T$ in the herding phase.
In the particular shown case, the herding pattern is in the``saving'' mode
and the distribution is skewed towards high $X_i$ values.
Not shown, is the corresponding distribution of the $S$-layer
which was in the herding ``pessimistic'' phase.}
\end{center}
\end{figure}

\begin{figure}[htp4]
\begin{center}
\includegraphics [angle=-90,scale=0.38]{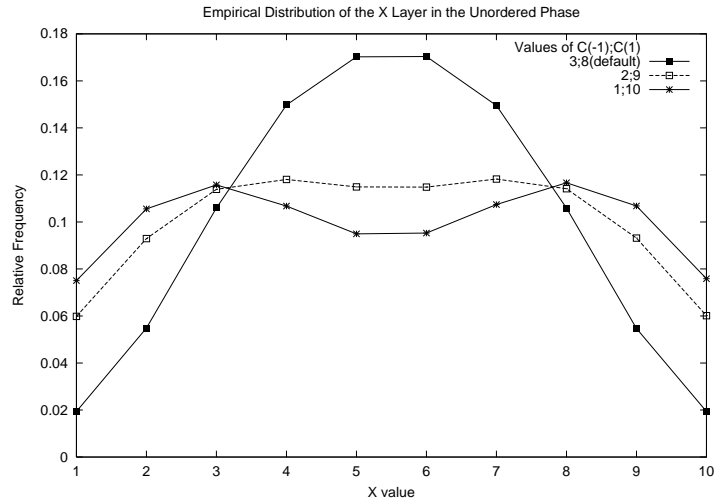}
\caption{Empirical distribution of individual savings ($X_i$) in
the unordered phase for different values of $C(\cdot)$. If $C(1)$
and $C(-1)$ are far enough from each other, a bimodal distribution
emerges, since most $X_i$ are in correlation with $S_{r(i)}$, and
the $S$ variables are distributed equally between 1 and -1 at the unordered phase.}
\end{center}
\end{figure}

\begin{figure}[htp6]
\begin{center}
\includegraphics[angle=-90,scale=0.4]{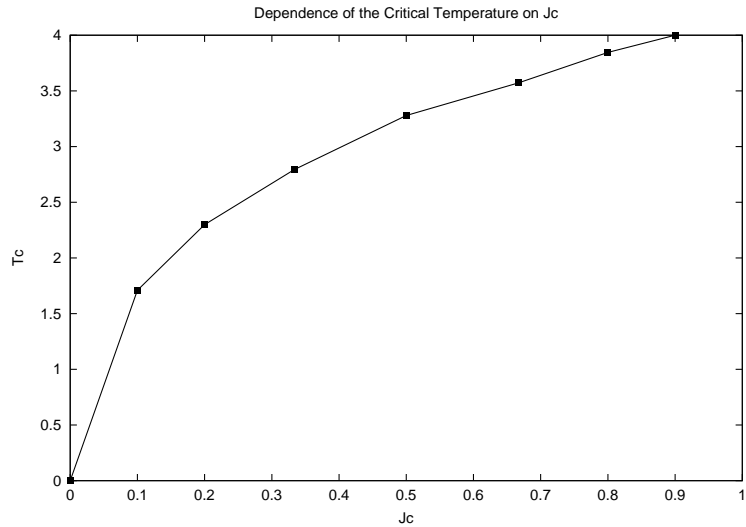}
\caption{Dependence of the $T_c$ on $J_c$. $J_x$
and $J_s$  are both held fixed at a value of $\frac{1}{3}$.
As $J_c$ vanishes the system effectively decouples into two independent
one-dimensional chains, and $T_c$ approaches $0$.}
\end{center}
\end{figure}

\begin{figure}[htp5]
\begin{center}
\includegraphics[angle=-90,scale=0.4]{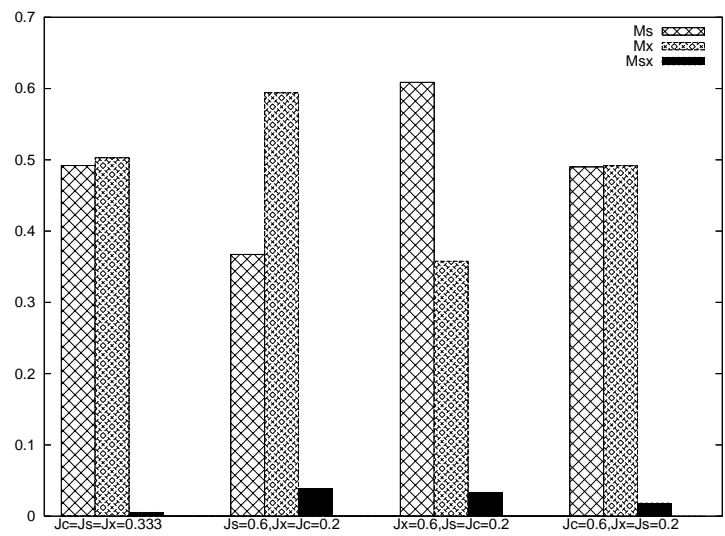}
\caption{Comparing ratios of various minority groups, 
out of the total number of agents not in the majority group, at different
values of coupling parameters. Note that for each set of bars, the ratios sum to 1.
The effect of increasing the
relative importance of one layer over the other is symmetric. 
When $J_x=J_s$, the minority groups are
symmetrical independently of $J_c$.}
\end{center}
\end{figure}
\end{document}